\documentclass[letterpaper]{article} 
\usepackage{aaai24}  
\usepackage{times}  
\usepackage{helvet}  
\usepackage{courier}  
\usepackage[hyphens]{url}  
\usepackage{graphicx} 
\def\UrlFont{\rm}  
\usepackage{natbib}  
\usepackage{caption} 
\usepackage{algorithm}
\usepackage{algorithmic}
\usepackage{booktabs}
\usepackage{amsmath}
\usepackage{amssymb}

\usepackage{newfloat}
\usepackage{listings}
\DeclareCaptionStyle{ruled}{labelfont=normalfont,labelsep=colon,strut=off} 
\floatstyle{ruled}
\newfloat{listing}{tb}{lst}{}
\floatname{listing}{Listing}
\title{Algorithm-Assisted Decision Making and Racial Disparities in Housing: A Study of the Allegheny Housing Assessment Tool}
\author{
   Lingwei Cheng\textsuperscript{\rm 1}, 
   Cameron Drayton\textsuperscript{\rm 1},
   Alexandra Chouldechova\textsuperscript{\rm 1},
   Rhema Vaithianathan\textsuperscript{\rm 2}
    \thanks
{We thank the Allegheny County Department of Human Services for their invaluable support and feedback.}
}
\affiliations{
    \textsuperscript{\rm 1}Heinz College of Information Systems and Public Policy, Carnegie Mellon University, Pittsburgh, PA,\\
    \textsuperscript{\rm 2}Auckland University of Technology, New Zealand\\
    \{lingweic, cdrayton, achoulde\}@andrew.cmu.edu, 
    rhema.vaithianathan@aut.ac.nz
}

\begin{document}

\maketitle




\begin{abstract} 
 The demand for housing assistance across the United States far exceeds the supply, leaving housing providers the task of prioritizing clients for receipt of this limited resource.  To be eligible for federal funding, local homelessness systems are required to implement assessment tools as part of their prioritization processes. The Vulnerability Index Service Prioritization Decision Assistance Tool (VI-SPDAT) is the most commonly used assessment tool nationwide.  Recent studies have criticized the VI-SPDAT as exhibiting racial bias \cite{courtney2022, Petry2021}, which may lead to unwarranted racial disparities in housing provision. In response to these criticisms, some jurisdictions have developed alternative tools, such as the Allegheny Housing Assessment (AHA), which uses algorithms to assess clients' risk levels. Drawing on data from its deployment, we conduct descriptive and quantitative analyses to evaluate whether replacing the VI-SPDAT with the AHA affects racial disparities in housing allocation. We find that the VI-SPDAT tended to assign higher risk scores to white clients and lower risk scores to Black clients, and that white clients were served at a higher rates pre-AHA deployment. While post-deployment service decisions became better aligned with the AHA score, and the distribution of AHA scores is similar across racial groups, we do not find evidence of a corresponding decrease in disparities in service rates. We attribute the persistent disparity to the use of Alt-AHA, a survey-based tool that is used in cases of low data quality, as well as group differences in eligibility-related factors, such as chronic homelessness and veteran status. We discuss the implications for housing service systems seeking to reduce racial disparities in their service delivery.
\end{abstract}

\section{Introduction}
On any given night, roughly half a million people in the United States are experiencing homelessness \cite{huduser2022ahar}.  Black and African American persons are disproportionately represented among the homeless population, with lifetime rates of homelessness estimated to be more than three times higher for non-Hispanic Blacks compared with non-Hispanic whites \cite{fusaro2018racial}. Some but not all of this disparity is explained by racial differences in income, incarceration rates, and exposure to traumatic events \cite{day2015racial}.   Providing long-term supportive housing is the single most effective intervention for reducing homelessness \cite{evans2019reducing}.

As a requirement for participating in programs funded by the United States Department of Housing and Urban Development (HUD), local homelessness systems must implement assessment tools that prioritize the most vulnerable homeless clients for supportive services. The most common housing assessment tool is an interview-based instrument called the Vulnerability Index Service Prioritization Decision Assistance Tool (VI-SPDAT) \cite{vi-spdat2016}. Problematically, recent studies suggest that the VI-SPDAT is poorly validated \cite{brown2018reliability} and racially biased \cite{courtney2022, Petry2021}. Partly in response to such criticism, homelessness systems across the country have been exploring alternative assessment tools, including ones based on predictive risk models.  

We study one such model developed and deployed by the Allegheny County Department of Human Services (ACDHS). Allegheny County is a medium-sized US county in Pennsylvania that contains the city of Pittsburgh. Allegheny County's Continuum of Care (CoC) receives HUD funding to coordinate and provide homelessness services such as shelters, supportive housing, outreach, engagement and assessment, and prevention strategies. ACDHS is responsible for managing HUD funding and operating the Coordinated Entry system that assesses and refers clients to CoC providers \cite{ac-continuum-charter}. 

In 2020, the county deployed a tool called the Allegheny Housing Assessment (AHA) to ``standardize the assessment process and move away from the interview-based VI-SPDAT approach to an automated assessment based on administrative data that can better predict the likelihood of future adverse outcomes'' \cite{vaithianathan_kithulgoda_2020}. In this paper, we study the transition from VI-SPDAT to the predictive risk model-based risk assessment, AHA, and evaluate the AHA tool's impact on racial disparities in access to supportive housing services. Specifically, we want to understand: 
\begin{enumerate}
    \item How did the prioritization for housing change when the VI-SPDAT was replaced with the AHA? 
    \item How does AHA impact the racial disparities in housing service enrollment rates? 
    \item How does the use of the alternative survey assessment tool (Alt-AHA) impact racial disparities? 
\end{enumerate}  

While there have been extensive evaluations of the performance, ethics, and algorithmic fairness of the AHA models \cite{vaithianathan_kithulgoda_2020, eticas-assessment, AHA-DHS-response}, we are among the first to study the impact of its deployment\footnote{We are aware of other ongoing work, but none that has been published.}.

Using extensive descriptive and quantitative analysis, we find that prior to the deployment of AHA, VI-SPDAT tended to assign higher risk scores to white clients and lower risk scores to Black clients. While there was no disparity in service rates conditioned on VI-SPDAT scores, the net effect was that white clients were served at a higher rate. Following the adoption of AHA, there was an increase in service to Black clients, especially those at medium- and high-risks. However, these increases paralleled those for white clients, and thus did not result in a reduction in racial disparities.  While there is no evidence of racial disparity among clients assessed with AHA only, there is a significant racial disparity in service rates among clients assessed with Alt-AHA.  White clients were about twice as likely as Black clients to be assessed with Alt-AHA, and conditional on being assessed were much more likely to be served.  The differences in Alt-AHA assessment rates is due in large part to differences in data quality, which is significantly lower for white clients.  Disparities in service rates across groups following an Alt-AHA assessment are at least partly attributable to other explanatory factors that affect eligibility, such as chronic homelessness, disability, and veteran status. 

Our study contributes to the literature on algorithmic fairness and human-AI interaction in several important ways. First, to our knowledge, it is the first study to investigate how algorithm-assisted decision-making affects housing disparities using real data from a deployed tool. Second, we directly study algorithm-assisted decision outcomes of an automated risk assessment tool in the presence of a manual assessment option (the Alt-AHA). Lastly, by identifying important differences between the widely used VI-SPDAT and a machine learning-based alternative (AHA), our study has important policy implications for housing providers.

\section{Background}
\label{section:background}
In response to longstanding housing shortfalls, in 2008 federal rules began incentivizing housing providers to use risk assessment tools to ensure that the most vulnerable clients are prioritized for supportive housing.  To receive federal funding, HUD requires housing providers to use tools that assess a client's housing and service needs and eligibility, and score their risk vulnerability \cite{hud2017cpdn}. 

\begin{figure}[t]
    \centering
    \includegraphics[scale=0.275]{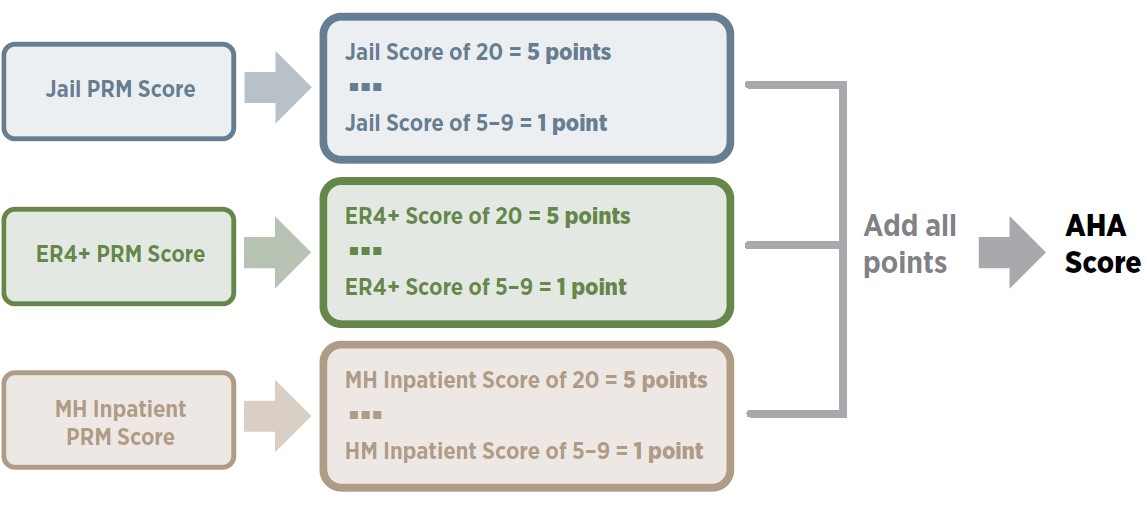}
    \caption{Mapping of Predictive Risk Model Scores to AHA Scores\citep{vaithianathan_kithulgoda_2020}}
    \label{fig:model_combo_rules}
\end{figure}

In this section we describe the most widely used tool, the VI-SPDAT, and then detail the development and deployment of the AHA tool that has since replaced the VI-SPDAT in ACDHS operations.

\subsection{VI-SPDAT}
The VI-SPDAT is a tool that combines OrgCode’s SPDAT (Service Prioritization Decision Assistance Tool) Prescreen tool and Community Solutions’ VI (Vulnerability Index). It features a survey for Coordinated Entry (CE) workers to assess client risk and need at intake. Gaining widespread adoption, it was used in 39 states and countries by 2015 \cite{nextGenTools} and remained in use in Allegheny County until 2020.

Recent research has concluded that the VI-SPDAT is poorly validated \cite{brown2018reliability} and racially biased \cite{courtney2022, Petry2021, caseworkerBias}. Moreover, the assessment questions can be invasive, asking about drug use, risk-taking behavior, and mental health diagnoses.  Focus groups conducted in Allegheny County showed that former clients had concerns about the long and detailed 45 minute interview the tool entailed, which specifically opened the potential to re-traumatize.  Despite the invasive questions, a 2018 analysis by ACDHS found that the VI-SPDAT was not effective at identifying clients who are at the greatest risk of adverse outcomes associated with housing instability, and the scores did not significantly predict risk of return to homeless services. Partly in response to the criticism, by December 2020 the maintainers of VI-SPDAT had announced that they would phase out the tool \cite{nextGenTools}. 

\subsection{The Allegheny Housing Assessment}
In 2014, ACDHS issued a request for proposals (RFP) to explore the use of predictive decision support tools in their operations.  The RFP resulted in a research partnership between ACDHS and the Center for Social Data Analytics at the Auckland University of Technology (CSDA). By 2018, CSDA began working on a machine learning--based risk assessment model that would eventually become the Allegheny Housing Assessment (AHA).

The AHA tool is based on three separate machine learning models, each trained to predict the likelihood of a given adverse event occurring in the 12 months following a housing services assessment: (i) a jail stay; (ii) a Medicaid-funded inpatient hospital stay for a behavioral health event; and (iii) 4 or more Medicaid-funded Emergency Room events.  Unlike the interview-based VI-SPDAT, the AHA relies not on self-reported answers to interview questions, but rather on demographic and county service utilization data stored in the Allegheny County Data Warehouse~\cite{alleghenydatawarehouse}.    Features considered by the models are drawn from child welfare services, county jail, courts, juvenile probation, behavioral and physical health, previous interactions with homeless services, previous interactions with assisted housing, demographics (other than race), neighborhood data from the American Community Survey, and household characteristics. 

Each of the three models produces a risk score between 1-20, and the three scores are then combined into a single AHA score. Figure~\ref{fig:model_combo_rules} below summarizes the combination step. Scores from the three separate models are binned, summed and then converted to deciles to produce a final AHA score between 1-10, with 10 being the highest risk. 

\subsection{The Alt-AHA Tool}
In certain cases, new clients may lack sufficient data to run the AHA tool. For this reason, ACDHS complements the AHA with Alt-AHA, an alternative risk assessment based on a subset of VI-SPDAT questions identified by tool developers as being  most predictive of risk. Alt-AHA was validated on the same outcomes as AHA, and calibrated to ensure that clients scoring 10 on Alt-AHA have a similar risk of adverse outcomes as those scoring 10 on AHA.

To help guide the use of Alt-AHA, each client on a referral receives a ``data quality indicator.''  The indicator is 1 if the Data Warehouse has a unique ID for the individual that has existed for at least 90 days and 0 otherwise. 

CE workers are required to use the Alt-AHA when: (i) the quality indicator is less than 1; or (ii) someone moved to Allegheny County within the last 90 days; or (iii)  someone has not consistently been living in Allegheny County; or (iv) someone is fleeing domestic violence and thus may have been unable to utilize services. Alt-AHA is sometimes used outside of these four cases. When CE workers use the Alt-AHA, they must record their reason for doing so.

An AHA score is always produced, irrespective of whether an Alt-AHA is conducted. The final score used in prioritizing clients, hereafter referred to as \textit{``posted risk score''}, is the greater of the AHA and Alt-AHA scores.  The next section describes how these scores are used as part of the broader intake process.  

\subsection{Housing Intake Process}
\label{section:intake_process}

\begin{figure}[t]
    \centering    \includegraphics[width=\linewidth]{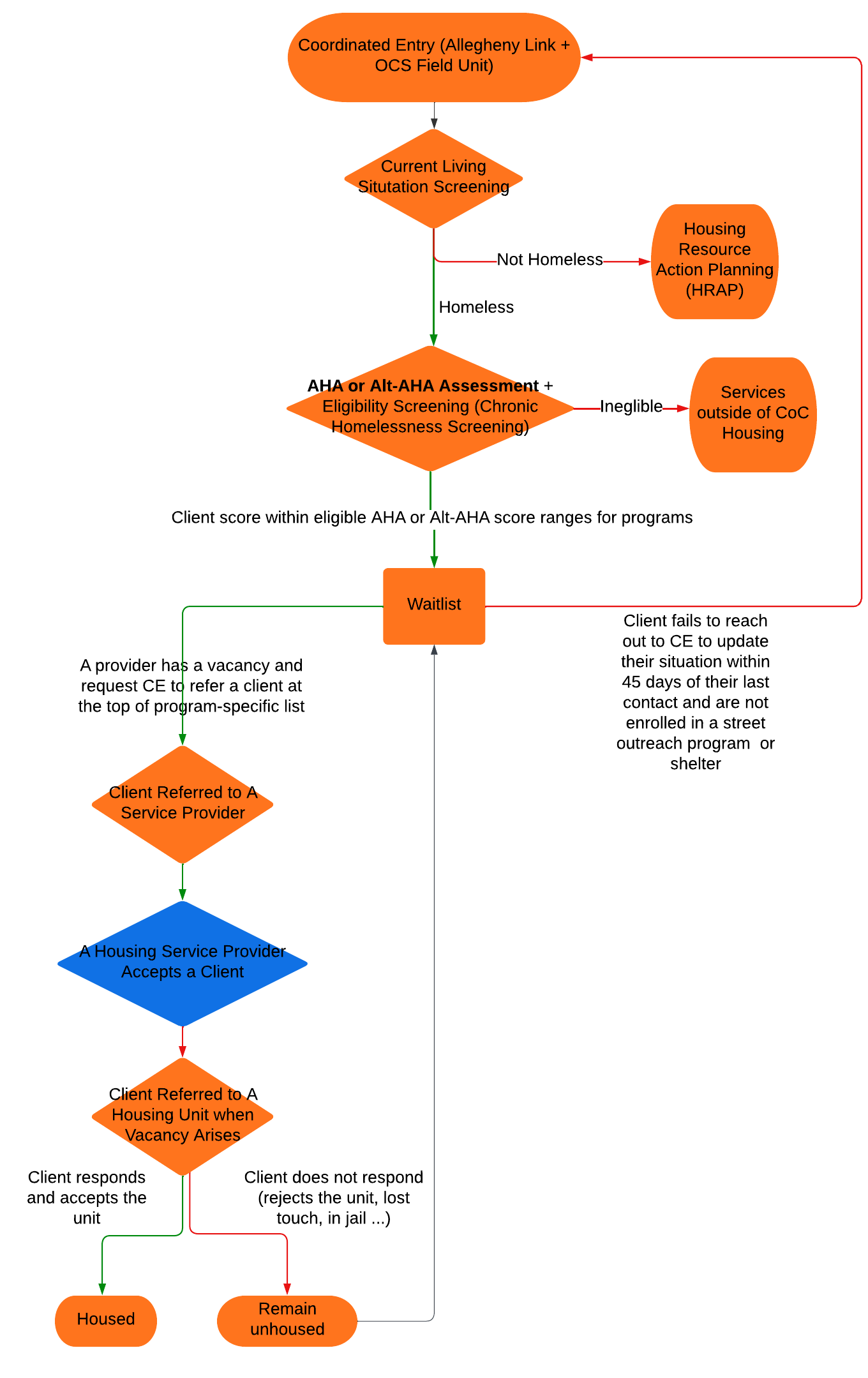}
    \caption{DHS Housing Services Intake Process}
    \label{fig:aha_process}
\end{figure}

The full client intake process is illustrated in Figure~\ref{fig:aha_process}. 
Clients can access the Coordinated Entry system for homeless services by contacting Allegheny Link or through referrals by the Office of Community Services (OCS) Field Units. For clients identified as homeless via the Current Living Situation Screening,  CE workers then conduct an assessment using AHA or Alt-AHA (prior to August 2020, VI-SPDAT).

After screening and scoring, CE workers assess clients' eligibility for specific housing programs. This entails screening for chronic homeless status, which can prioritize them for housing \cite{hud-chronically-homeless-finalrule}. The combination of (Alt-)AHA scores and special circumstances (e.g. veteran status, fleeing from domestic abuse, or disability status) dictates program eligibility. The CE coordinator then places eligible clients on waitlists for the appropriate housing types. If the CE cannot reach a client for updates every 45 days from their last contact, and the client is not enrolled in a street outreach program or shelter, they are removed from the waitlist.

Once a provider has an available slot and the capacity to serve and find a client housing, they request a referral from ACDHS off the waitlist. The housing provider then enrolls clients into their programs, offering case management services and housing assistance. Importantly, this does not guarantee housing; clients might not accept the offer due to personal preferences or circumstances such as hospitalization or incarceration. Despite the AHA tool's key role in housing allocation, human factors like maintaining the waitlist and locating clients still play a crucial part in the process, thus influencing when clients are referred. However, CE workers' ability to alter a client's waitlist position is limited by strict eligibility rules. This is in contrast to other algorithm-assisted decision-making settings such as bail or child maltreatment screening, where human decision-makers have considerably more discretion. 

\section{Related Literature}
\label{section:literature}
Algorithm-assisted decision-making is increasingly used in the public sector, including in homelessness prevention.  Predictive models utilize proxy outcomes, administrative data, and target specific sub-populations to efficiently allocate limited resources in crucial settings. Various studies have used these models to predict outcomes such as high-cost users \cite{Toros}, first-time homelessness \cite{von2019predicting}, returns to homelessness \cite{von2019predicting, Kube_Das_Fowler_2019}, shelter entries \cite{shinn2013efficient}, and chronic homelessness \cite{messier2022predicting, vanberlo2021interpretable}. 
Additional research focuses on specific groups, predicting risks such as daily sheltering arrangements for homeless youth \cite{ijerph17186873} and their susceptibility to substance abuse \cite{tabar20, Dou_Barman-Adhikari_Fang_Yadav_2021}, as well as assessing homelessness risks among veterans \cite{KOH202213, Byrne2019}. 
The AHA tool differs from these other approaches by focusing on long-term risks associated with homelessness, providing algorithmic scores rather than binary risk assessments to prioritize clients on housing waitlists managed by ACDHS.

Recent studies on predictive tools in housing service allocation reveals several significant challenges. The first concern is the accuracy of administrative data in reflecting clients' past circumstances and urgent needs \cite{Sheng_Shen23, karusala19}. Research by \citet{karusala19} involving interviews with policymakers and caseworkers shows mixed feelings about the objectivity and accuracy of the VI-SPDAT. Factors like clients' trust in caseworkers and their privacy preferences impact what they disclose, which in turn affects scoring.  The scoring also depends on the coordinators' ability to prompt questions and gauge the effects of a client's mood and mental status. Thus, the fairness of these tools may suffer from discrepancies in what is intended to be measured versus what is actually being measured and assessed in the AI model \cite{jacobs18}.   \citet{karusala19} note that tools like AHA, which rely on administrative rather than self-reported data, address mismeasurement attributable to faulty self-reporting, but remain susceptible to other biases in data collection.

The second concern focuses on the use of proxy outcomes in prediction, which may not fully capture the subjective experiences of unhoused individuals \citet{Sheng_Shen23}. Research shows that ML systems might de-prioritize crucial values like efficiency and privacy that are essential for protecting homeless individuals \cite{dilruba2023}. Building trust in these systems at provider and bureaucratic levels is crucial for their support, as noted by \citet{saxena21}.

Debate around fairness in housing services also persists. \citet{Kube_Das_Fowler_2019} highlight a trade-off between fairness and efficiency, where imposing fairness constraints can increase re-entry rates.  \citet{karusala19} note a division between service providers, who prioritize vulnerability over race, and policymakers, who seek to address racial disparities in housing access. Some researchers advocate for using reparative algorithms to rectify historical inequities \cite{so2022}. Our study specifically evaluates racial disparities in the allocation of housing.

ACDHS commissioned an independent equity evaluation \cite{eticas-assessment} of the models prior to deployment, which found that the models operated consistently across racial sub-groups.

\section{Methodology}
\label{section:methodology}

\subsection{Research Questions}
\label{section:research_question}
In the study, we seek to understand how CE workers interact with the AHA tool and how the AHA tool impacts racial disparity in housing allocation.   More specifically, we ask three Research Questions (RQ's): 
\begin{enumerate}
    \item How did the prioritization for housing change when the VI-SPDAT was replaced with the AHA? 
    \item How does AHA impact the racial disparities in housing service enrollment rates? 
    \item How does the use of the alternative survey assessment tool (Alt-AHA) impact racial disparities? 
\end{enumerate}  

We study changes in \textit{service rates}.  An assessment is ``served'' if it results in an enrollment with a Permanent Supportive Housing (PSH), Rapid Rehousing (RR), or Transitional Housing (TH) provider.  This means the client can start receiving comprehensive case management services, such as mental health and substance use treatment, and housing assistance. As discussed in \textsection~\ref{section:intake_process}, enrollment does not mean they are assigned to or have moved into housing yet. We do not distinguish between housing types, as all three provide services that address housing instability barriers.
This study is not a comprehensive evaluation of the impact of the tool on overall welfare---only on racial disparities. 

\subsection{Descriptive Analysis and Empirical Model}
We conduct descriptive analyses to examine high-level trends, and then more rigorously study the research questions using logistic regressions to control for client characteristics. Table~\ref{table:notation} provides a summary of key variables and how they are denoted throughout the paper. We also discuss the distinction between demographics and housing eligibility characteristics in Appendix~\ref{section:appendix_demographics}.

To answer RQ1, we use retrospectively generated AHA scores for the pre- and post-deployment periods to investigate changes in service rates.  We conduct the following regression on clients who were assessed only via AHA, focusing on $\beta_3$.  
\begin{gather}
\small
\begin{aligned}
    Served_{i,t} & = \beta_0 + \beta_1 Post_{i,t} + \beta_2 Score  + \beta_3 Score*Post_{i,t}
    \\& + \beta_4 Black_{i,t} + \beta_5'X + \gamma_t + \delta_t + \epsilon_{i,t} 
\end{aligned}
\label{eq:rq1}
\end{gather}
$\beta_3$ captures the change in association between service decisions and AHA scores from pre to post. We control for client race ($Black$ indicator variable), as well as other controls, $X$, which include age, gender, veteran status, chronic homelessness status, disability status, etc., as shown in Table~\ref{table:notation}. Time-fixed effects $\gamma_t$ and $\delta_t$ represent assessment year and month, respectively, and capture yearly and seasonal variations in service provision. We use robust standard errors, clustered by client ID, to account for multiple assessments per client and for among-client variation.  

We study RQ2 with the following specification, using data only on clients where AHA (not Alt-AHA) was used.   
\begin{gather}
\small
\begin{aligned}
    Served_{i,t} & = \beta_0 + \beta_1 Post_{i,t} + \beta_2 Black_{i,t} + \beta_3 {Score}_{i,t} \\
    & + \beta_4 {Risk Group}_{i,t} + \beta_5 Black * Post_{i,t} \\
    & + \beta_6 {Risk Group} * Post_{i,t} + \beta_7 {Risk Group} * Black_{i, t} \\
    &  + \beta_8 {Risk Group} * Black * Post_{i,t} \\
    & + \beta_9' X + \gamma_t + \delta_t + \epsilon_{i,t} 
\label{eq:rq2}
\end{aligned}
\end{gather}
This model allows for heterogeneity in the association between service decisions, race, and AHA deployment across risk group strata. The risk strata are defined in Table~\ref{table:notation} as low-risk (AHA scores 1-4), medium-risk (scores 5-8), and high-risk (scores 9-10). We present a simplified short-hand equation here and include an expanded equation in Appendix~\ref{section:appendix_research_methods} for clarification. 

Two coefficients of interest, $\beta_5$ and $\beta_8$, measure changes in service rates. $\beta_5$ reflects how service rates change for the low-risk group post-AHA. $\beta_8$ shows the additional service rate change for Black clients in the high-risk group, compared to white clients, pre- and post-AHA deployment, relative to the low-risk group. Essentially, it shows the impact of high risk status on racial disparities post-AHA. As a robustness check, we test alternative specifications in Appendix~\ref{section:appendix_robustness_check}, treating AHA scores as continuous or binary for high-risk. The results are consistent with the main findings.

For RQ3, we use logistic regression on data from clients assessed with Alt-AHA, using the following specification where the Alt-AHA score, if higher, is used as the posted score for housing prioritization.
\begin{gather}
\small
\begin{aligned}
    Served_{i,t} & = \beta_0 + \beta_1 Black_{i,t} + \beta_2 {Posted Score}_{i,t} \\ 
    & + \beta_3{Use Alt Score} + \beta_4{Quality Indicator} \\ 
    & + \beta_5' X + \gamma_t + \delta_t + \epsilon_{i,t} 
\end{aligned}
\label{eq:rq3}
\end{gather}
\verb|QualityIndicator| assesses data quality on a 0 to 1 scale. All other covariates $X$ and time fixed effects remain consistent. The coefficient of interest, $\beta_1$, evaluates potential racial disparities in service rates for clients prioritized based on Alt-AHA after controlling for other factors.

\section{Data}
\label{section:data_descriptive}
We extract housing assessment data from the County’s Homelessness Management Information System (HMIS) for assessments conducted between January 1, 2018 and March 29, 2022. This data is linked with enrollment data for clients in PSH, RR, or TH services during the study period. 

We retrospectively (re-)run the AHA tool to produce scores for pre- and post-AHA deployment periods. We also obtain VI-SPDAT scores from the pre-deployment period. 

We exclude assessments with a race other than white or Black/African American, as they represent only 3.2\% of the data. We also exclude 36.1\% of the assessments associated with families or youth household types, focusing instead on single adult household types. Prioritization of families or youth households is more complicated and depends on a range of additional criteria not captured by our data. Lastly, we exclude clients who are enrolled with programs that do not require AHA scores. The final dataset contains 6,542 assessments, in which 66.1\% were from the pre-AHA period. 399 Alt-AHA assessments were conducted, with 370 using the Alt-AHA scores as their posted scores.

Table~\ref{table:descriptive} summarizes pre- and post-deployment client characteristics.  We see that the majority of single homeless clients are male, and over 90\% report a disability. Post deployment, the prevalence of chronically homeless clients rises. When comparing Black and white clients, their assessment scores, ages, and likelihood of being domestic violence survivors remain similar pre- and post-deployment. However, they diverge in key eligibility criteria: white clients exhibit higher rates of chronic homelessness and disability, whereas Black clients are more likely to be veterans.

Before deployment, Black clients have a lower average VI-SPDAT score (7.5) than white clients (8.4) on a 1-16 scale. Post-deployment, Black clients average a slightly higher AHA score (7.1) compared to white clients (6.6) on a 1-10 scale. White clients are twice (24.5\%) as likely to be assessed with Alt-AHA compared to Black clients (11.8\%), and to have those scores used for prioritization (20.7\% vs. 8.0\% for Black clients). This significant difference may be linked to consistently higher data quality indicators for Black clients, with scores of 0.96 vs. 0.87 pre-deployment and 0.91 vs. 0.82 post-deployment.

Finally, white clients were served at a higher rate than Black clients before AHA deployment, 17.6\% compared to 14.5\%.  The disparity persists post-deployment.

\begin{table}[t]
\centering
\scalebox{0.85}
{%
\begin{tabular}{lcccc}
\toprule
& \multicolumn{2}{c}{Pre-deployment} & \multicolumn{2}{c}{Post-deployment} \\
\cmidrule(lr){2-3} \cmidrule(lr){4-5}
 & Black/AA & White & Black/AA & White \\
\midrule
\% Female               & 33.1 & 32.1 & 35.1 & 37.3 \\
                        & (47.1) & (46.7) & (47.7) & (48.4) \\
\% Male                 & 59.9 & 62.1 & 64.9 & 62.6 \\
                        & (49.0) & (48.5) & (47.8) & (48.4) \\
\% Unknown G       & 7.0  & 5.8  & 0.1  & 0.1  \\
                        & (25.5) & (23.3) & (3.0) & (3.1) \\
Age                     & 46.3 & 45.7 & 46.1 & 44.0 \\
                        & (12.3) & (12.0) & (12.2) & (11.7) \\
\%  CH     & 25.2 & 29.8 & 35.4 & 38.6 \\
                        & (43.4) & (45.8) & (47.8) & (48.7) \\
\% Disability           & 91.2 & 95.2 & 91.0 & 93.9 \\
                        & (28.4) & (21.5) & (28.7) & (24.0) \\
\% Veteran              & 9.1  & 7.8  & 7.8  & 6.4  \\
                        & (28.8) & (26.8) & (26.8) & (24.5) \\
\% DV Survivor & 11.5 & 11.4 & 13.5 & 13.6 \\
                        & (32.0) & (31.8) & (34.2) & (34.3) \\
AHA Score               & 6.5  & 6.3  & 7.1  & 6.6  \\
                        & (2.7) & (2.7) & (2.5) & (2.8) \\
VI-SPDAT Score          & 7.5  & 8.4  & {-} & {-} \\
                        & (2.5) & (3.1) & {-} & {-} \\
Alt-AHA Score           & {-} & {-} & 5.1  & 6.1  \\
                        & {-} & {-} & (2.7) & (2.5) \\
\% Use Alt-AHA    & {-} & {-} & 11.8 & 24.5 \\
                        & {-} & {-} & (32.3) & (43.0) \\
\% Use Alt Score        & {-} & {-} & 8.0  & 20.7 \\
                        & {-} & {-} & (27.1) & (40.5) \\
Quality Indicator & 0.96 & 0.87 & 0.91 & 0.82 \\
                      & (0.20) & (0.32) & (0.28) & (0.38) \\
Served                  & 14.5 & 17.6 & 19.5 & 23.3 \\
                        & (35.2) & (38.1) & (39.7) & (42.3) \\
N                       & {2,176} & {2,147} & {1,141} & {1,078} \\
\bottomrule
\multicolumn{5}{l}{\footnotesize Standard deviation in parentheses}\\
\multicolumn{5}{l}{\footnotesize Unknown G = Unknown Gender, CH = Chronic Homeless}\\
\multicolumn{5}{l}{\footnotesize DV = Domestic Violence}\\
\end{tabular}
}
\caption{Summary Statistics by Race and Deployment Status}
\label{table:descriptive}
\end{table}

\section{Descriptive Analysis}
We now compare risk scores and service rates from VI-SPDAT, AHA, and Alt-AHA to gain high-level insight relating to our research questions.  In Appendix~\ref{section:appendix_concurrent_events} we provide a brief analysis of concurrent events, particularly the impact of COVID-19 and lockdown orders, surrounding the AHA deployment. While we note some variation in service rates by race, these changes are not statistically significant.

\subsection{Score Comparisons}
\label{section:aha_vs_vispdat}
\subsubsection{AHA vs. VI-SPDAT Scores}
\begin{figure}[t]
    \centering
    \includegraphics[width=\linewidth]{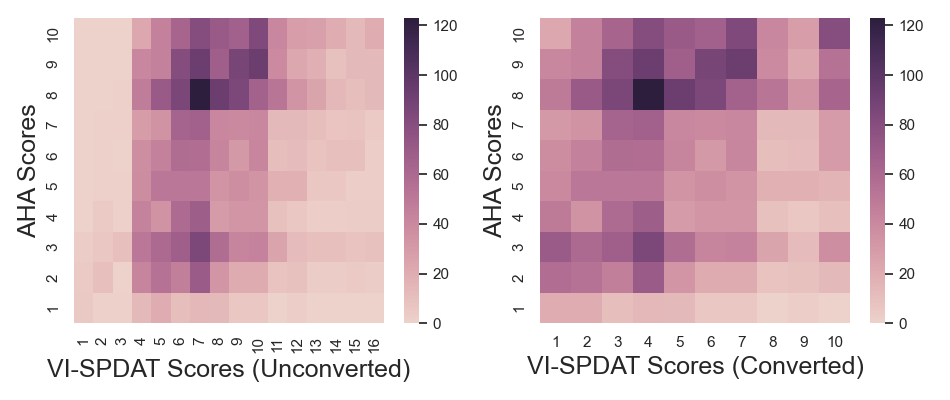}
    \caption{AHA vs. VI-SPDAT Scores Prior to AHA Implementation. Original VI-SPDAT scores (left), converted VI-SPDAT scores (right).
    }
    \label{fig:aha_spdat_correlation}
\end{figure}

We begin by comparing the AHA and VI-SPDAT scores. Because the VI-SPDAT was no longer conducted after the AHA was deployed, this comparison is based on data during the pre-deployment period.  VI-SPDAT scores range from 1-16 where 1 is the lowest and 16 is the highest risk, and AHA scores range from 1-10.  The left panel on Figure~\ref{fig:aha_spdat_correlation} shows that there is a weak positive correlation between the retrospectively generated AHA scores and the unconverted VI-SPDAT scores (Spearman's \( \rho = 0.218\), \( p < 0.0001 \)).

To facilitate a more direct comparison of the scores for the remainder of the analysis, we bin the VI-SPDAT scores into a 1-10 range.  Scores 1-4 are mapped to 1, 13-16 are mapped to 10, and values in-between are mapped in a one-to-one correspondence to 2:9.  This mapping is monotonic, and is chosen to ensure that there are comparable numbers of clients in the lowest and highest score bins for both the AHA and (post-mapping) VI-SPDAT.   Figure~\ref{fig:aha_spdat_correlation} (right) shows a heatmap of scores after mapping. After conversion, the AHA and VI-SPDAT Scores remain significantly weakly positively correlated (Spearman's \( \rho = 0.217\), \( p < 0.0001 \)) 

\subsubsection{AHA vs. Alt-AHA Scores} 
\begin{figure}[h]
    \centering
    \includegraphics[width=0.6\linewidth]{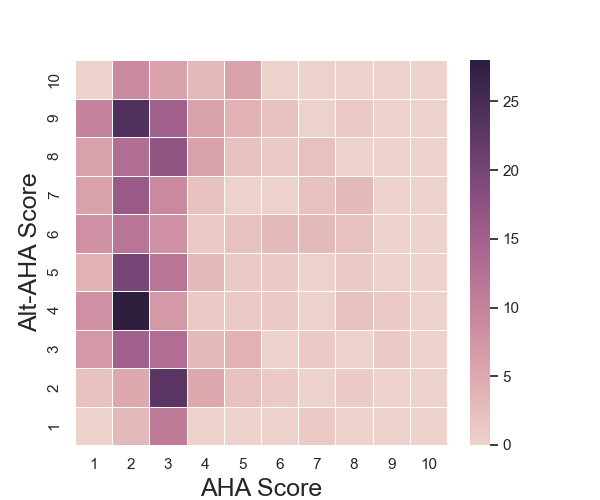}
    \caption{AHA vs. Alt-AHA Scores for clients who were assessed with Alt AHA. Clients with low AHA scores (1-3) typically received higher Alt-AHA scores.}
    \label{fig:aha_alt_aha_correlation}
\end{figure}

Figure~\ref{fig:aha_alt_aha_correlation} compares AHA and Alt-AHA scores for clients assessed with Alt-AHA post-deployment. Clients with initial low AHA scores (1-3) typically received higher Alt-AHA scores. As a business rule, the \textit{higher} of the two scores is used as the score (``posted score'') for prioritization. Alt-AHA and AHA scores do not appear correlated (Spearman's \( \rho = -0.02\), \( p = 0.64 \)).

\begin{figure}[t]
    \centering
    \includegraphics[width = 1\linewidth]{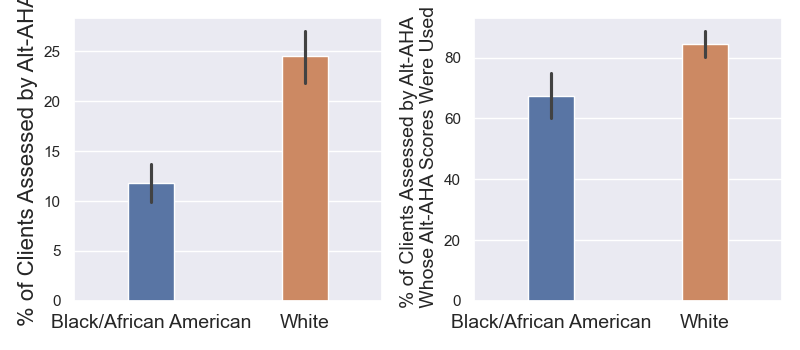}
    \caption{Percent of Clients Assessed by Alt-AHA by Race (left); and Percent of Clients Assessed by Alt-AHA whose Alt-AHA Scores were used for Prioritization by Race (right). White clients are twice as likely to be assessed via Alt-AHA than Black clients. Of those assessed by Alt-AHA, white clients' Alt-AHA scores are also more likely to be used for prioritization.
    }
    \label{fig:aha_alt_aha_use_rate}
\end{figure}

Figure~\ref{fig:aha_alt_aha_use_rate} shows a breakdown of Alt-AHA use by race.  We observe that white clients are more likely than Black clients to be assessed via Alt-AHA (left panel), and also more likely to then have their Alt-AHA scores used (right panel). These differences are at least partly attributed to lower quality indicators for white clients, as shown in Table~\ref{table:descriptive}. CE workers provide reasons for disagreeing with AHA scores, which we discuss in Appendix~\ref{section:appendix_score_disagree_reasons}.  There we find that white clients are more often not residing consistently in Allegheny County, have lower data quality indicators, and have lower service utilization than Black clients. 

\subsection{Score Distribution by Race}
\label{section:score_distribution_service_rate}
\begin{figure*}[t]
    \centering
    \includegraphics[width=1\linewidth]{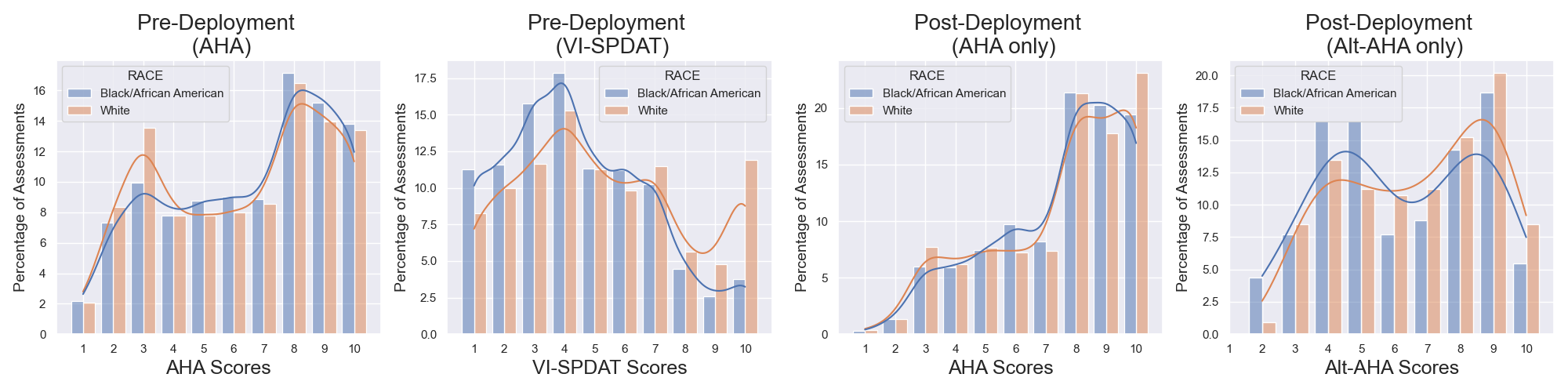}
    \caption{AHA and VI-SPDAT Score Distributions by Deployment Period and Race, Normalized by 
 Race. Normalization within each racial group shows the proportion of individuals receiving each risk score. Pre-AHA deployment, VI-SPDAT classified more Black clients as low-risk and more white clients as high-risk, while AHA would have classified more white clients as low-risk. Post-AHA deployment, the score distribution shifted towards higher risk levels. Both AHA and Alt-AHA scored Black and white clients at comparable rates.}
    \label{fig:score_distributions}
\end{figure*}

Figure~\ref{fig:score_distributions} shows the (frequency) distribution of pre-deployment VI-SPDAT and AHA scores, and post-deployment AHA and Alt-AHA scores\footnote{A cumulative density version of these plots is shown in Figure~\ref{fig:score_cumulative_distribution_above} of Appendix~\ref{section:appendix_score_distribution}.}.  We present within-racial-group frequencies instead of counts so that pre-vs-post comparisons are not confounded by variation in the number of clients being assessed over time.

Comparing the pre-deployment plots, we see that VI-SPDAT scores were lower for Black clients than white clients, with white clients being more likely to receive VI-SPDAT scores in the 8-10 range.  Conditional on VI-SPDAT score, Figure~\ref{fig:score_service_rate} shows that Black and white clients were equally likely to receive housing services.  However, at least in part because VI-SPDAT scores tended to be lower for Black clients, overall service rates were significantly lower for Black clients than white clients pre-AHA (see Figure~\ref{fig:service_rate_bars}). 

Unlike the VI-SPDAT, in both the pre- and post-deployment period, Black clients were approximately equally likely to receive AHA scores in the 8-10 range.   The third panel of Figure~\ref{fig:score_distributions} shows the subsample where AHA scores determined prioritization---this combines clients assessed only via AHA and those assessed via Alt-AHA but whose AHA score was higher. Here, Black and white clients exhibit similar score distributions.  To the extent that differences in prioritization score distributions drive disparities in service rates, this analysis indicates that disparities might decrease once AHA is introduced.  

Second, transitioning from VI-SPDAT to AHA, we observe a significant shift in the score distribution, with a larger percentage of clients across both races now classified as higher risk, whereas previously, most fell into the low to medium risk category. Post-AHA deployment, there is an increased representation of Black clients in the medium and high-risk categories, compared to the VI-SPDAT. 

Finally, in the case of clients assessed via Alt-AHA whose Alt-AHA scores were used for prioritization (where the posted scores are identical to their Alt-AHA scores), the distribution shows a slightly higher percentage of Black clients in the lower risk range and a slightly lower percentage in the high risk range compared to white clients. Given the small sample size, the difference between the distributions is unlikely to be statistically significant.

In summary, we observe a notable disparity under VI-SPDAT, with a higher percentage of Black clients categorized as lower risk and white clients as higher risk. Post-deployment, both Black and white clients receive similar scores under AHA, with Black clients showing an increased presence in the medium to high-risk categories compared to their earlier VI-SPDAT scores. White clients tend to receive higher risk scores when assessed by Alt-AHA. 

\subsection{Service Rates by Race}
\label{section:service_rate_by_race}
\begin{figure*}[t]
    \centering
    \includegraphics[width = 0.75\linewidth]{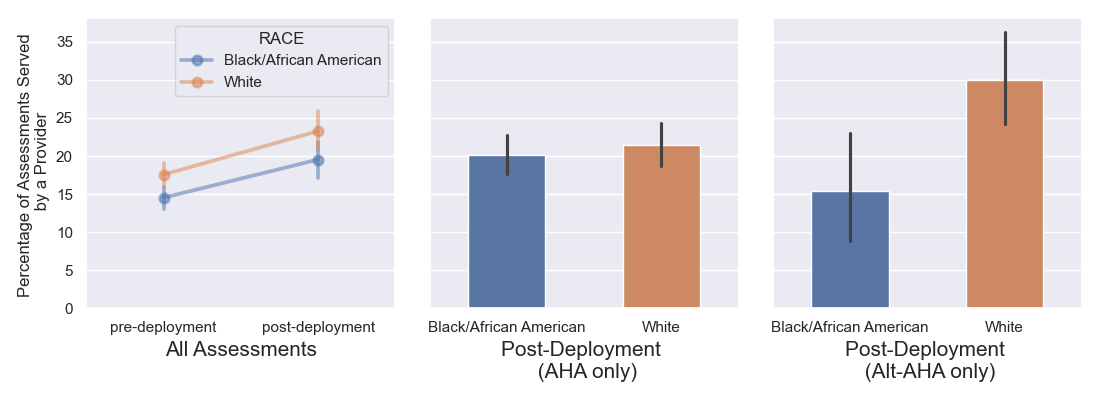}
    \caption{Percent of Assessments Served by a Provider by Race Pre- and Post-AHA deployment. Pre-deployment, providers served white clients at a higher rate than Black clients. Post-deployment, service rates increased for both, but white clients  continued to be served at higher rates. AHA-assessed clients are served at similar rates by race. White Alt-AHA assessed clients are served at significantly higher rates.}
    \label{fig:service_rate_bars}
\end{figure*}

According to Figure~\ref{fig:service_rate_bars}, pre-AHA, white clients were served at a significantly higher rate than Black clients  ($t\text{-statistic}=-2.72$, $\text{P-value}=0.006$). Post-deployment, the service rates increased for both groups, but white clients continued to be served at a significantly higher rate than Black clients ($t\text{-statistic}=-2.15$, $\text{P-value}=0.03$). When we break down post-deployment assessments by whether the Alt-AHA scores were used, we find that the two groups are served at similar rate when AHA scores were used. The racial disparity in service rates appears to be driven mainly by the sub-group of assessments where an Alt-AHA was used. For clients assessed via Alt-AHA, on aggregate, there is a statistically significantly difference ($t\text{-statistic}=-2.72$, $\text{P-value}=0.007$) between Black and white clients, with white clients (30.0\%) about twice more likely to be enrolled with a provider than Black clients (15.4\%). However, it is important to note that the Alt-AHA sample size is small, and the aggregate comparison overlooks risk scores and other client eligibility factors like veteran status and chronic homelessness. Also, data quality for white clients is generally lower with Alt-AHA. We will revisit this more rigorously in \textsection~\ref{section:results_rq3}.

\begin{figure*}[t]
    \centering
    \includegraphics[width = 0.95\linewidth]{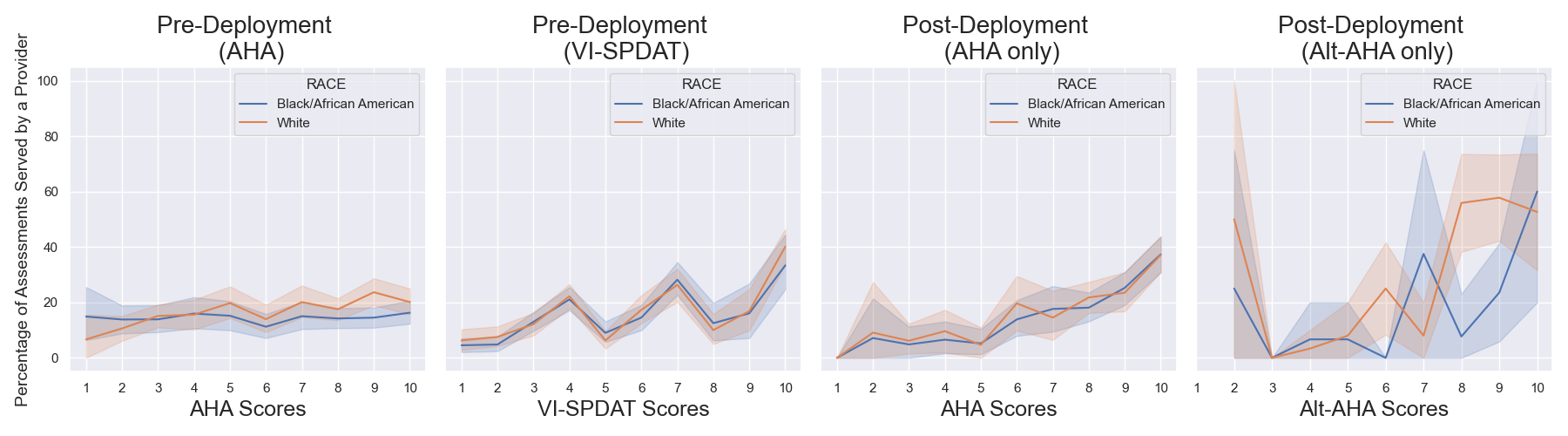}
    \caption{Percent of Assessments Served by Race for Pre- and Post-deployment Periods, Conditioned on Scores. }
    \label{fig:score_service_rate}
\end{figure*}

Figure~\ref{fig:score_service_rate} shows service rates conditioned on AHA and VI-SPDAT scores.  We note that in the pre-AHA period, VI-SPDAT scores are strongly correlated with the likelihood of enrolling with a provider. In contrast, conditioned on the retroactively calculated AHA risk scores, the service rates appear uncorrelated with AHA scores.  Post-deployment, we observe that service rates aligned more closely with AHA scores, with higher scores more likely to be served.  

Furthermore, conditioned on VI-SPDAT scores, we observe no racial disparities in service rates during the pre-deployment period. However, as mentioned previously, this lack of association does not imply an absence of racial disparity, since VI-SPDAT tended to underestimate the risk of Black clients. If we consider the proxy harms that AHA measures, when CE workers assessed risks using VI-SPDAT without access to AHA scores in the pre-deployment period, white clients were served at higher rates than Black clients among those whose AHA risk score was 9 or 10, as shown in the first panel to the left in Figure~\ref{fig:score_service_rate}. Post-deployment, the introduction of AHA scores appears to have closed this disparity for clients in the high-risk category. We will test this hypothesis more rigorously in \textsection~\ref{section:results_rq2}.
 
When examining the third and fourth panels in Figure~\ref{fig:score_service_rate}, which depict service rates for AHA and Alt-AHA assessments, respectively, we observe no significant gap in service for assessments conducted via AHA. While there is a significant overall difference between Black and white clients assessed via Alt-AHA, it disappears when conditioned on Alt-AHA scores---likely due to small sample sizes.

\subsection{Conclusions of Descriptive Analysis}
Addressing RQ1, prior to AHA deployment, service rates are highly associated with VI-SPDAT scores, and largely uncorrelated with AHA scores. However, post-AHA deployment, service rates begin to correlate strongly with AHA scores, especially for clients assessed exclusively with AHA.

Addressing RQ2, both pre- and post-deployment, white clients were served at a significantly higher rate than Black clients. While there is no overall disparity in service rates in clients assessed by AHA only, service rates for clients assessed by Alt-AHA are higher for white clients than Black clients.  There is no racial disparity conditioned on VI-SPDAT scores pre-AHA. Yet, retrospective AHA scoring identify two key service rate disparities.  (1) VI-SPDAT tended to assign higher risk scores to white clients compared to Black clients.  By contrast,  AHA scores are more comparable across groups.  (2) Under VI-SPDAT, white clients were served at significantly higher rates than Black clients at the same retrospective AHA risk scores, particularly in the high-risk category. The deployment of AHA reduced the service rate disparity among high-risk clients. 

Finally, for RQ3, we find that post-AHA deployment, white clients were more likely to be assessed via Alt-AHA, their Alt-AHA scores were more likely to be higher than their AHA scores compared to Black clients, and they were more likely to be served. However, this may be due to differences in data quality, demographics, and eligibility criteria. 

\section{Quantitative Results}
In this section, we further interrogate the findings from the descriptive analysis using logistic regressions to control for additional explanatory variables. 

\subsection{RQ1}
\label{section:results_rq1}
Because we retrospectively (re-)calculated both pre- and post-deployment AHA scores, we can examine whether the switch from VI-SPDAT to AHA resulted in service decisions becoming more closely aligned with AHA scores. 

Table~\ref{table:rq1} shows the results of testing this hypothesis using regression specification~\eqref{eq:rq1}.  In the table columns we progressively control for demographics and eligibility criteria, and then also time fixed effects.

\begin{table}[t]
\centering
\def\sym#1{\ifmmode^{#1}\else\(^{#1}\)\fi}
\small
{
\begin{tabular}{l*{3}{c}}
\hline\hline
            &\multicolumn{1}{c}{(1)}&\multicolumn{1}{c}{(2)}&\multicolumn{1}{c}{(3)}\\
\hline
Post        &     -2.01\sym{***}&     -2.01\sym{***}&     -2.24\sym{***}\\
            &    (0.33)         &    (0.42)         &    (0.46)         \\
[0.25em]
AHA Score       &      0.05\sym{**} &     -0.02         &     -0.01         \\
            &    (0.02)         &    (0.02)         &    (0.02)         \\
[0.25em]
AHA Score * Post  &      0.28\sym{***}&      0.31\sym{***}&      0.31\sym{***}\\
            &    (0.04)         &    (0.04)         &    (0.04)         \\
[0.25em]
Black       &     -0.19\sym{**} &     -0.15         &     -0.16\sym{*}  \\
            &    (0.07)         &    (0.08)         &    (0.08)         \\
[0.25em]
D \& E Crit. &                      & $\checkmark$         & $\checkmark$     \\
[0.25em]
Quarter FE  &                      &                      & $\checkmark$     \\
[0.25em]
Yearly FE   &                      &                      & $\checkmark$     \\
\hline
N           &   6,143         &   6,143         &  6,143         \\
Pseudo. R\(^2\)       &      0.03         &      0.28         &      0.28         \\
\hline\hline
\multicolumn{4}{l}{\footnotesize Standard errors in parentheses}\\
\multicolumn{4}{l}{\footnotesize \sym{*} \(p<0.05\), \sym{**} \(p<0.01\), \sym{***} \(p<0.001\)}\\
\end{tabular}
}
\caption{Service Rate by AHA Implementation}
\label{table:rq1}
\end{table}

Comparing the \verb|AHA Score| coefficient across columns (1) and (2), we see that once eligibility criteria are controlled for, there's no statistically significant association between decisions and the AHA score \textit{pre-}deployment.  On the other hand, the coefficient of \verb|AHA Score * Post| is consistently statistically significant across columns, indicating that service rates became significantly positively associated AHA scores \textit{post}-deployment. Across all columns, the coefficients on \verb|Black| consistently show that Black clients are statistically significantly less likely to be served compared to white clients.

\subsection{RQ2}
\label{section:results_rq2}
As discussed in \textsection~\ref{section:score_distribution_service_rate}, when CE workers relied solely on VI-SPDAT without access to AHA scores, they tended to serve white clients at a higher rate within the AHA score range 9 and above. We further examine the change in service rate by risk groups using the Equation~\ref{eq:rq2} allowing for heterogeneous treatment effect. We present the results in Table~\ref{table:triple-diff-diff}. 

\begin{table}[t]\centering
\def\sym#1{\ifmmode^{#1}\else\(^{#1}\)\fi}
\scalebox{0.85}
{
\def\sym#1{\ifmmode^{#1}\else\(^{#1}\)\fi}
\begin{tabular}{l*{3}{c}}
\hline\hline
            &\multicolumn{1}{c}{(1)}&\multicolumn{1}{c}{(2)}&\multicolumn{1}{c}{(3)}\\
\hline
Black       & 0.09         & 0.15         & 0.16         \\
            & (0.16)       & (0.19)       & (0.19)       \\
[0.25em]
Post        & -0.63        & -0.66        & -0.83        \\
            & (0.36)       & (0.46)       & (0.49)       \\
[0.25em]
AHA Score   & 0.12\sym{**} & 0.06         & 0.06         \\
            & (0.04)       & (0.05)       & (0.05)       \\
[0.25em]
High Risk   & -0.22        & -0.26        & -0.25        \\
            & (0.30)       & (0.35)       & (0.35)       \\
[0.25em]
Medium Risk & -0.16        & -0.06        & -0.05        \\
            & (0.22)       & (0.25)       & (0.25)       \\
[0.25em]
High Risk * Black & -0.52\sym{*} & -0.47 & -0.51 \\
            & (0.21)       & (0.26)       & (0.26)       \\
[0.25em]
Medium Risk * Black & -0.37 & -0.42 & -0.42 \\
            & (0.20)       & (0.24)       & (0.24)       \\
[0.25em]
High Risk * Post & 1.12\sym{**} & 1.41\sym{**} & 1.34\sym{**} \\
            & (0.39)       & (0.46)       & (0.46)       \\
[0.25em]
Medium Risk * Post & 0.54 & 0.72 & 0.65 \\
            & (0.39)       & (0.46)       & (0.46)       \\
[0.25em]
Black * Post & -0.46 & -0.54 & -0.59 \\
            & (0.54)       & (0.64)       & (0.64)       \\
[0.25em]
High Risk * Black * Post & 0.87 & 0.85 & 0.93 \\
            & (0.58)       & (0.69)       & (0.69)       \\
[0.25em]
Medium Risk * Black * Post & 0.60 & 0.62 & 0.65 \\
            & (0.59)       & (0.69)       & (0.70)       \\
[0.25em]
D \& E Crit. &                     &    $\checkmark$   &   $\checkmark$    \\
[0.25em]
Quarter FE      &            &              &  $\checkmark$      \\
[0.25em]
Yearly FE                &                     &        &      $\checkmark$   
\\
\hline
N           &   6,143         &   6,143         &   6,143        \\
Pseudo. R\(^2\)        &      0.03         &      0.28         &      0.28        \\
\hline\hline
\multicolumn{4}{l}{\footnotesize Standard errors in parentheses}\\
\multicolumn{4}{l}{\footnotesize \sym{*} \(p<0.05\), \sym{**} \(p<0.01\), \sym{***} \(p<0.001\)}\\
\end{tabular}
}
\caption{Service Rate by Race Risk Group, and AHA Implementation}
\label{table:triple-diff-diff}
\end{table}

We find that that the coefficients of \verb|High Risk * Post| are consistently statistically significant across the specifications, indicating that high risk white clients are more likely to be served post-AHA deployment compared to pre. The coefficients of \verb|High Risk * Black * Post|  and \verb|Medium Risk * Black * Post| are not statistically significant for either specification that controls for demographics and eligibility criteria.  We therefore do not have sufficient evidence to conclude that the change in service rates post-deployment is different for Black clients than for white clients for any of the risk strata.  
    
\subsection{RQ3}
\label{section:results_rq3}
As for the use of Alt-AHA, across all models shown in Table~\ref{table:rq3}, higher posted scores and quality indicator scores are consistently associated with a higher likelihood of receiving services through Alt-AHA usage. Conversely, the use of Alt-AHA alone does not significantly impact service likelihood. 

When considering the effect of race and posted score (the higher of Alt-AHA or AHA scores that is used for prioritization for this sub-population) in column (1) and (2) and (3) on service rates, Black clients are significantly less likely to receive services through Alt-AHA compared to white clients, even after adjusting for differences in data quality (in (3)). Once demographic, eligibility and time fixed effects are added to the regression, the race coefficient is no longer statistically significant.  However, this is more due to an increase in the standard error of the coefficient estimate than a clear decrease in the magnitude of the coefficient.  The evidence of higher service rates for white clients under Alt-AHA is therefore borderline.  

\begin{table}[t]\centering
\def\sym#1{\ifmmode^{#1}\else\(^{#1}\)\fi}
\scalebox{0.85}
{
\begin{tabular}{l*{5}{c}}
\hline\hline
            &\multicolumn{1}{c}{(1)}&\multicolumn{1}{c}{(2)}&\multicolumn{1}{c}{(3)}&\multicolumn{1}{c}{(4)}&\multicolumn{1}{c}{(5)}\\
\hline
Black  &     -0.79\sym{**} &     -0.61\sym{*}  &     -0.69\sym{*}  &     -0.57         &     -0.66         \\
            &    (0.28)         &    (0.31)         &    (0.32)         &    (0.37)         &    (0.38)         \\
[0.25em]
Posted Score   &                     &      0.48\sym{***}&      0.50\sym{***}&      0.50\sym{***}&      0.54\sym{***}\\
            &                     &    (0.07)         &    (0.08)         &    (0.10)         &    (0.12)         \\
[0.25em]
Use Alt AHA &                     &                     &     -0.08         &     -0.30         &     -0.45         \\
            &                     &                     &    (0.40)         &    (0.52)         &    (0.53)         \\
[0.25em]
QI &                     &                     &      1.04\sym{***}&      0.84\sym{*}  &      0.77\sym{*}  \\
            &                     &                     &    (0.28)         &    (0.36)         &    (0.36)         \\
[0.25em]
D \& E Crit. &            &     &               & $\checkmark$         & $\checkmark$     \\
[0.25em]
Quarter FE  &          &    &                &                    & $\checkmark$     \\
[0.25em]
Yearly FE   &         &    &                 &                      & $\checkmark$     \\
\hline
N           &    399         &    399         &    399         &    371         &    371         \\
Pseudo. R\(^2\)       &      0.02         &      0.18         &      0.21         &      0.41         &      0.42         \\
\hline\hline
\multicolumn{5}{l}{\footnotesize Standard errors in parentheses}\\
\multicolumn{5}{l}{\footnotesize \sym{*} \(p<0.05\), \sym{**} \(p<0.01\), \sym{***} \(p<0.001\)}\\
\multicolumn{5}{l}{\footnotesize QI = Quality Indicator.}\\
\multicolumn{5}{l}{\footnotesize disability != 1 predicts failure perfectly; }\\
\multicolumn{5}{l}{\footnotesize disability omitted and 28 obs not used.}\\
\end{tabular}
}
\caption{Alt-AHA Service Rate}
\label{table:rq3}
\end{table}

\subsection{Conclusion of Quantitative Analysis}
In summary, our quantitative analysis confirms, as we would expect, that post-AHA deployment, service decisions become more closely aligned with AHA scores. Overall, we do not find evidence that the AHA deployment significantly bridges the disparity in service rates between Black and white clients. The post-deployment change in service rates for medium and high risk Black clients is similar to that observed for white clients.  Lastly, while there is borderline evidence that white clients are more likely to be served when Alt-AHA is used, ultimately the sample size underpinning that analysis is too small to draw strong conclusions.

\section{Limitations}
Despite modeling AHA risk scores within actual decision-making contexts, the complex journey from initial assessment to final service enrollment requires us to make certain assumptions in our interpretation.

First, it is essential to acknowledge the possibility that the effects identified in our study are in part attributable to changes during the COVID period related to housing policy and client composition that influenced service allocations.  We further discuss the observed effects of the stay-at-home order and other COVID-era impacts in Appendix \textsection\ref{section:appendix_concurrent_events}. 

In order to conclude that the effects we identified are attributable to the deployment of AHA rather than other unmeasured factors, we need to assume that there are no ``unobservables'' that both (i) changed at the same time as AHA was deployed, and (ii) were relied upon by CE workers in their service decisions. It could be problematic for our study design if, for instance, some information gathered during the assessment process but not reflected in the administrative data influenced CE worker decisions.  We were not able to identify alternative observational study designs that would not be susceptible to potential unobserved confounding.

\section{Discussion}
We present the first evaluation of the impact of replacing the VI-SPDAT with the Allegheny Housing Assessment on housing service outcomes.  Specifically, we examine how CE decisions change post-deployment, and the effects of those changes on racial disparities in service rates.  While earlier studies of AHA considered the predictive performance and group-fairness properties of the tool's constituent models, they relied on pre-deployment data and did not study the tool's influence on decisions.

We find that white clients were significantly more likely than Black clients to receive high VI-SPDAT scores, but were approximately equally likely to receive high AHA scores.  Later analyses do not, however, enable us to conclude that the shift to AHA-based prioritization significantly reduced the racial disparities in service decisions present under the VI-SPDAT policies.  Pre-AHA, white clients were served at higher rates than Black clients, both overall and conditional on AHA score.  We find that racial disparities largely persisted post-AHA, particularly in the subset of clients assessed with Alt-AHA.  Our analysis indicates that the persistent disparities are at least in part attributable to differences in data quality, and in differences in eligibility criteria across groups. 

This study has several policy implications for CoC providers considering a switch from the VI-SPDAT to predictive models like the AHA tool or other survey-based assessment tools such as the Alt-AHA. Our results show that predictive models like AHA \textit{can} reduce or altogether eliminate differences in the prioritization score distributions.  This, however, \textit{does not} guarantee that overall disparities in service decisions will be reduced as a result.  Whereas we found no significant post-deployment disparity in service rates among clients assessed with AHA, disparities among clients assessed with Alt-AHA were at least as large as those existing pre-deployment.  This finding underscores difficulties with anticipating the impact of a predictive risk model deployment in settings where the model cannot be run due to poor data quality in a substantial number of cases. 

As discussed extensively in the literature review, there is an important concern about whether administrative data can accurately capture clients' past and current needs \cite{Sheng_Shen23}. Efforts to improve data accuracy and enhance cultural competence among service providers are critical for addressing disparities effectively. Policymakers must consider the strengths and weaknesses of risk assessments derived from administrative data and predictive modeling. Effectively integrating AI with CE workers' expertise could be pivotal in reducing racial disparities in housing service provision.

Future research should examine how CE workers utilize Alt-AHA or diverge from standard AHA procedures to support their clients. Such insights could clarify additional risks identified during interviews and guide improvements in algorithm iterations, enhancing decision-making in social services.

\section*{Researcher Positionality Statement}
All authors have previously conducted evaluations of algorithmic tools in either child welfare or housing services deployed by the ACDHS for academic research. The second author has interned part-time at the ACDHS, gaining firsthand experience with their operational processes and tools. The last author led the development of the AHA tool, and the third author provided feedback on the methodology used in model development and validation.  As a result, the research team collectively has a deep understanding of the design, development, validation and deployment process surrounding the AHA tool.  

We recognize that this experience and familiarity also present the potential for conflicts of interest. In conducting the research described in this paper we maintained a firm commitment to academic integrity and transparency.  As per the terms of our data use agreement, ACDHS was provided with the submission version of this manuscript and given the opportunity to review and provide feedback.  The data use agreement does not enable ACDHS to block the publication.  Through this review process we received helpful feedback that helped improve the clarity of the manuscript, led to a better-motivated model specification for RQ1, and also resulted in several robustness checks now presented in the appendix.  

Our collective expertise includes backgrounds in public policy, statistics, and health economics, enabling a multidisciplinary approach to this evaluation. We are motivated by a shared goal of improving the efficacy and fairness of algorithmic tools in social services.  Our work is informed not only through collaborations with government agencies, but also through engagement with stakeholders most likely to be impacted by the technologies.


\bibliography{main-AIES}

\appendix

\section{Impact of Concurrent Events During Tool Launch on Racial Disparity}
\label{section:appendix_concurrent_events}
\begin{figure*}[h]
    \centering
    \includegraphics[scale=0.2]{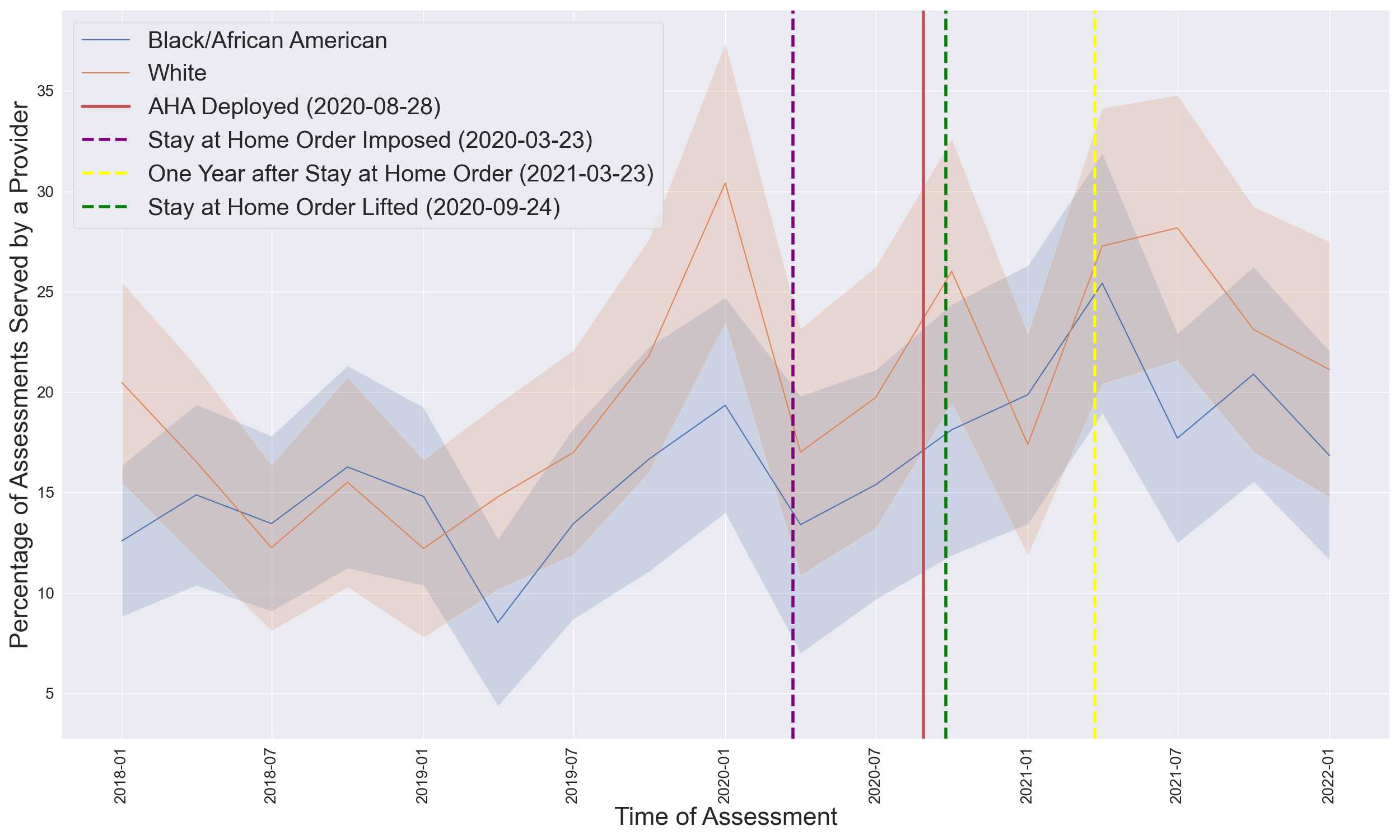}
    \caption{Service Rate by Race Overtime by Quarter. We mark three important events in our analysis period that prompted ACDHS to adjust their programs and could potentially affect racial disparity in service allocations: Pennsylvania's "Stay at Home Order" issued by the state's Department of Health on March 23rd, 2020, the lift of the order in Allegheny County on September 24th, 2020, and one year post the start of the order. Despite the order's implementation, we observe no significant statistical difference in service rates between Black and white clients over time.}
    \label{fig:overtime_service_rate}
\end{figure*}

Figure~\ref{fig:overtime_service_rate} shows the service rate for each quarter by race over our analysis period. We highlight three events within our analysis period that could affect service rates and racial disparity: the issuance of the "Stay at Home Order" by Pennsylvania and the state's Department of Health on March 23rd, 2020 \cite{pittsburgh-city-paper}, the lift of the order in Allegheny County on September 24th, 2020 \cite{cbs-pittsburgh-2020}, and one year post the start of the order. These dates are denoted by the purple, green, and yellow dashed lines, respectively, on Figure~\ref{fig:overtime_service_rate}. In response to the order and HUD COVID-requirements, ACDHS had to swiftly modify their programs, specifically reducing the number of unhoused clients. During the implementation of the order, we observe increasing service rates for both Black and white clients, aligned with ACDHS's change in business practice. After the AHA deployment, followed shortly by the lift of the order, we observe a short period of fall in the white-Black disparity - but the disparity grew again in favor of white clients one year post after the stay-at-home order. However, it's important to note that these differences are unconditioned on risk levels and also lack statistical significance, a point we assess more rigorously in \textsection~\ref{section:results_rq2}.

\section{Reasons for Using Alt-AHA}
\begin{figure*}[h]
    \centering
    \includegraphics[scale=0.4]{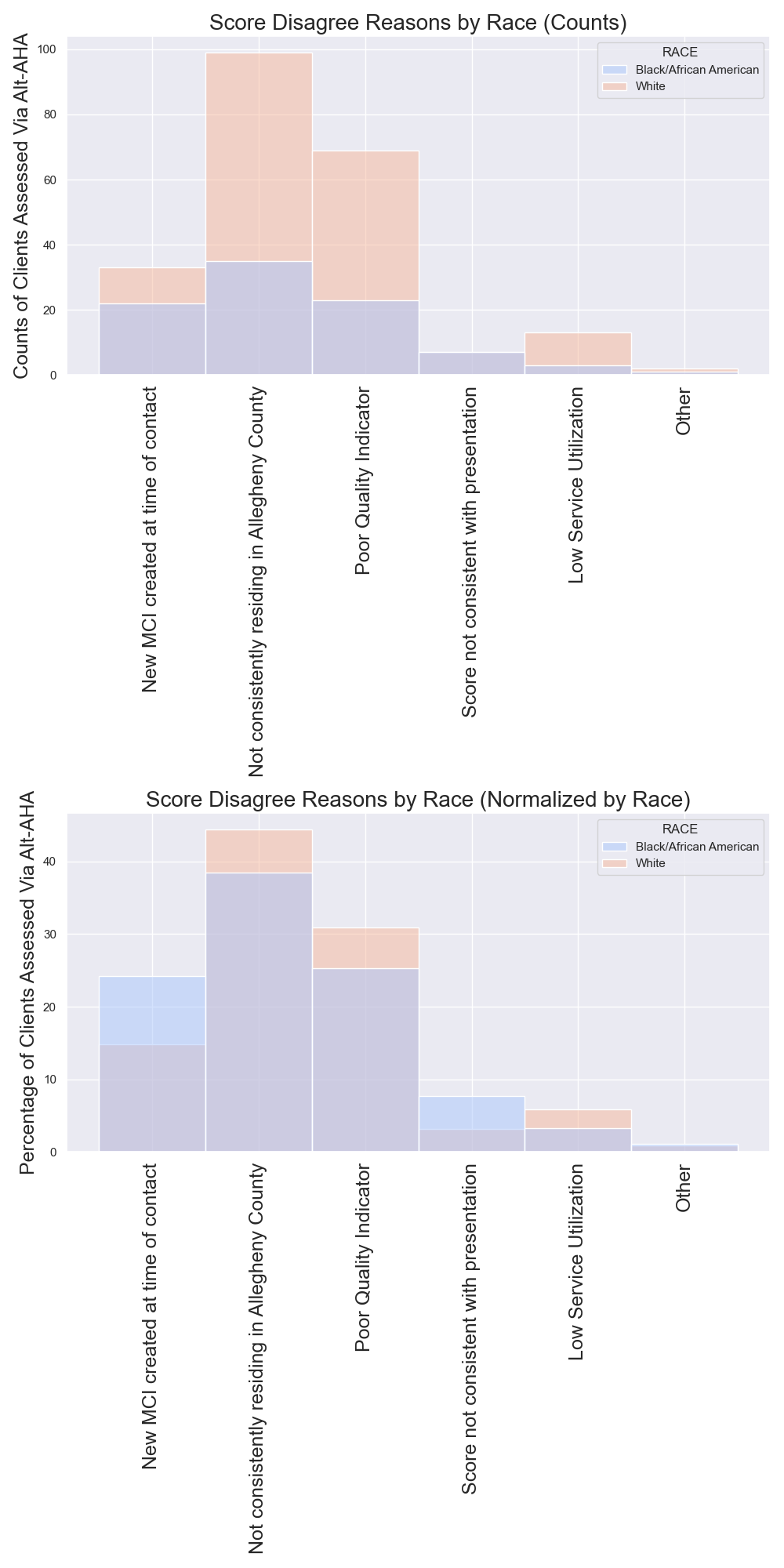}
    \caption{Scores Disagree Reasons by Race (Counts and Percentages)}
    \label{fig:score_disagree_reasons}
\end{figure*}

In Figure~\ref{fig:score_disagree_reasons} we show the reasons for why CE workers disagreed with the AHA scores and decided to conduct Alt-AHA instead. We show the counts of the reasons by race and the percentages of the reasons within each race respectively in the top and bottom graphs. We see that CE workers disagreed with AHA scores for white clients mostly because they do not live consistently in Allegheny county, have poor quality indicators and have low service utilization.  CE workers disagreed with AHA scores for Black clients mostly because they do not live consistently in Allegheny county, have poor quality indicators and did not have client ID (MCI) in the system at the time of contact. 

\label{section:appendix_score_disagree_reasons}

\section{Score Distribution}
\label{section:appendix_score_distribution}

\begin{figure*}[h]
    \centering
    \includegraphics[scale=0.35]{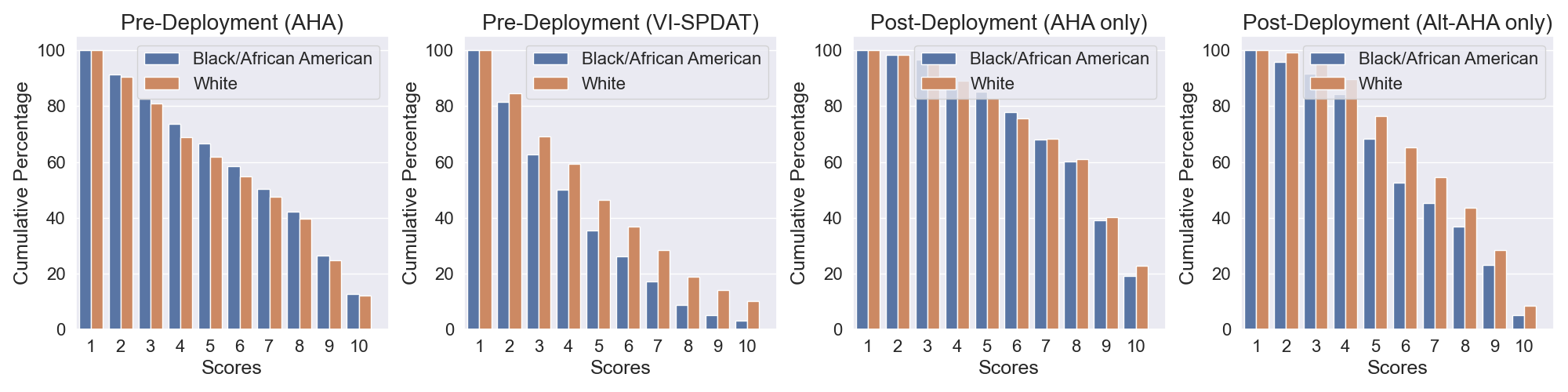}
    \caption{Cumulative Distribution of AHA and VI-SPDAT Scores by Deployment Phase and Race (Above Threshold). This figure shows the cumulative percentage of assessments above each score by race. AHA scores are more comparable between Black and white clients pre- and post-deployment, while VI-SPDAT consistently scores a higher percentage of white clients above each threshold.}
\label{fig:score_cumulative_distribution_above}
\end{figure*}

In Figure~\ref{fig:score_cumulative_distribution_above}, we show the percentage of assessments above each scores by race. All assessments has a score one or above, therefore we start with 100\% for both white and Black clients. The size of the bars diminishes as there are smaller percentages of higher risk scores. We can interpret the graphs as showing the percentage of clients above a score threshold, where the higher the score the more vulnerable and therefore more eligible for services. We see that both pre- and post-AHA deployment, AHA generates more comparable scores for both racial groups. In contrast, at every scores, there are consistently higher percentages of white clients than Black clients according to VI-SPDAT. With similar service rates under VI-SPDAT, this highlights the scoring mechanism as a potential source of disparity as there are always a higher percentage of white clients above every score thresholds. 

\section{Research Methods}
\label{section:appendix_research_methods}
Here we expand the equation in described in \textsection~\ref{section:methodology}. low-risk is 1 if a client's AHA score is between 1-4 and 0 otherwise. medium-risk is 1 if a client's AHA score is between 5-8 and 0 otherwise. high-risk is 1 if a client's AHA score is between 9-10 and 0 otherwise. We chose low-risk group as our base group for comparison. Thus, the coefficient of interest here is $7H$ which shows the additional difference in change in service rate for Black clients in the high-risk group pre and post-AHA deployment compared to white clients compared to such change in the low-risk group. 

\begin{align}
    Served_{i,t} &= \beta_0 + \beta_1 Post_{i,t} + \beta_2 Black_{i,t} \nonumber \\
    &\quad + \beta_3 Score_{i,t} + \beta_{4L} Low Risk_{i,t} \nonumber \\
    &\quad + \beta_{4M} Medium Risk_{i,t} + \beta_{4H} High Risk_{i,t} \nonumber \\
    &\quad + \beta_5 (Black \times Post_{i,t}) \nonumber \\
    &\quad + \beta_{6L} (Low Risk_{i,t} \times Post_{i,t}) \nonumber \\
    &\quad + \beta_{6M} (Medium Risk_{i,t} \times Post_{i,t}) \nonumber \\
    &\quad + \beta_{6H} (High Risk_{i,t} \times Post_{i,t}) \nonumber \\
    &\quad + \beta_{7L} (Low Risk_{i,t} \times Black_{i, t}) \nonumber \\
    &\quad + \beta_{7M} (Medium Risk_{i,t} \times Black_{i, t}) \nonumber \\
    &\quad + \beta_{7H} (High Risk_{i,t} \times Black_{i, t}) \nonumber \\
    &\quad + \beta_{8L} (Low Risk_{i,t} \times Black \times Post_{i,t}) \nonumber \\
    &\quad + \beta_{8M} (Medium Risk_{i,t} \times Black \times Post_{i,t}) \nonumber \\
    &\quad + \beta_{8H} (High Risk_{i,t} \times Black \times Post_{i,t}) \nonumber \\
    &\quad + \beta_9 X + \beta_{10} AltAHA + \gamma_{t} + \delta_{t} + \epsilon_{i,t} 
\end{align}

\begin{table*}[t]
\centering
\small{
\begin{tabular}{@{}lp{3.5cm}p{8cm}@{}}
\toprule
Term                        & Notation   & Description     
\\ \midrule
\textbf{Dependent Variable}  &            &                                                                                                                                                                                                                 \\
Served                      & $Served \in \{0,1\}$          & 1 if a client is enrolled by a housing service provider, 0 otherwise                                                                                                                                                                       \\
\textbf{Independent Variables}        &            &                                                                                                                                                                                                                 \\
AHA Score                       & $AHA Score \in \{1,\ldots, 10\}$          & An integer AHA risk score between 1 and 10 assessed by the AHA tool.                                                                                                                                              \\
Posted Score                 & $Posted Score \in \{1,\ldots, 10\}$         & An integer risk score between 1 and 10 that is the maximum (AHA Score, Alt-AHA Score) for assessments where Alt-AHA were conducted                                                        \\
Use Alt Score                        &  $Use Alt Score \in \{0, 1\}$& 1 if the Alt-AHA scores were used for prioritization, 0 otherwise (only applicable to assessments where Alt-AHA were conducted)                                                        \\
Quality Indicator                        &  $Quality Indicator \in \{0, \ldots, 1\}$& a continuous variable showing the quality of the client's data with 0 being the lowest and 1 being the highest.                                                       \\
Black                       & $Black \in \{0, 1\}$& 1 if the client is Black, 0 otherwise                                                           \\
Post                        & $Post \in \{0, 1\}$& 1 if the assessment was taken post AHA implementation, 0 otherwise                                                           \\
Risk Group                  & $RiskGroup \in \{\text{high, medium, low}\}$ & A categorical variable indicating if a client's AHA score is high risk (9-10), medium (5-8), or low (1-4)                        \\
High Risk                   & $HighRisk \in \{0, 1\}$ & 1 if the client's AHA score is 9 or 10, 0 otherwise                                     \\
Medium Risk                 & $MediumRisk \in \{0, 1\}$ & 1 if the client's AHA score is between 5 and 8 (inclusive), 0 otherwise                        \\
Low Risk                    & $LowRisk \in \{0, 1\}$ & 1 if the client's AHA score is between 1 and 4 (inclusive), 0 otherwise                    \\
Assessment Characteristics  & $X$          & A vector of variables on the characteristics of an assessment including the client's age, gender, veteran status, chronic homeless status, disability status, whether a client is a domestic violence survivor, whether the client is the head of the household, the size of the household, the type of household (Single or Households without Children).\\
Yearly Fixed Effects        & $\gamma$          & Yearly fixed effects controlling for the yearly variation in service rate                                                                                                                                 \\
Quarterly Fixed Effects     & $\delta$          & Quarterly fixed effects controlling for seasonal variation in service rate               \\            
\bottomrule
\end{tabular}
}
\caption{Terminology and Notation}
\label{table:notation}
\end{table*}

\section{The Effects of Demographics and Eligibility Criteria}
\label{section:appendix_demographics}

According to the AHA Methodology Update report \cite{AHA_Methodology_Update}, demographics include only age and gender—race is not considered when generating AHA scores. Therefore, we classify age and gender as demographics, while categorizing all other features, such as chronic homelessness status, disability, veteran status, and domestic violence survivor status, as eligibility criteria based on discussions with ACDHS (see section~\ref{section:intake_process} for details about the business process).

In Table~\ref{table:demographics_on_score}, we use OLD regression, and regress demographics and a combination of demographics and eligibility criteria on AHA scores using the sub-sample for which Alt-AHA was never conducted, indicating sufficient data quality for AHA scores. This allows us to disentangle the effects of demographics alone and demographics combined with eligibility criteria on AHA scores.

Unsurprisingly, demographics alone explain only 1.92\% of the variation in scores, as indicated by the adjusted R-squared value, because demographic features are just a small subset of all features used for prediction. However, after including eligibility criteria, we find that race, demographics, and eligibility criteria together explain 10.70\% of the variation in scores, suggesting that eligibility criteria are significantly associated with the scores. Despite this, the explanatory power remains relatively low. Thus, while adding eligibility criteria is expected to reduce the size of the coefficients associated with AHA scores due to collinearity, they will also directly impact service rates.

\begin{table*}[t]
\centering
\scalebox{0.85}{
{
\def\sym#1{\ifmmode^{#1}\else\(^{#1}\)\fi}
\begin{tabular}{l*{2}{c}}
\hline\hline
            &\multicolumn{1}{c}{(1)}&\multicolumn{1}{c}{(2)}\\
\hline
Black  &      0.2148\sym{*}  &      0.2624\sym{**} \\
            &    (0.0867)         &    (0.0831)         \\
[1em]
Gender Female   &     -0.2710         &     -0.4576\sym{***}\\
            &    (0.2140)         &    (0.0962)         \\
[1em]
Gender Unknown    &      -         &     -0.0132         \\
            &         (-)         &    (0.2074)         \\
[1em]
Gender Male    &      0.2195         &      -         \\
            &    (0.2047)         &         (-)         \\
[1em]
Age         &     -0.0246\sym{***}&     -0.0236\sym{***}\\
            &    (0.0035)         &    (0.0034)         \\
[1em]
Chronic Homeless&                     &      0.6375\sym{***}\\
            &                     &    (0.0762)         \\
[1em]
Disability  &                     &      1.8576\sym{***}\\
            &                     &    (0.1186)         \\
[1em]
Is Veteran  &                     &     -0.5812\sym{***}\\
            &                     &    (0.1615)         \\
[1em]
Is Domestic Violence Survivor&                     &     -0.1046         \\
            &                     &    (0.1180)         \\
[1em]
Household Size&                     &     -0.3052         \\
            &                     &    (0.2347)         \\
[1em]
Single Household&                     &     -0.9896\sym{***}\\
            &                     &    (0.0735)         \\
[1em]
Head of Household&                     &      0.9292\sym{*}  \\
            &                     &    (0.3720)         \\
\hline
N           &   6,143         &   6,143         \\
Adjusted R\(^2\)
&        0.0192                       &              0.1070          \\
\hline\hline
\multicolumn{3}{l}{\footnotesize Standard errors in parentheses}\\
\multicolumn{3}{l}{\footnotesize \sym{*} \(p<0.05\), \sym{**} \(p<0.01\), \sym{***} \(p<0.001\)}\\
\end{tabular}
}
}
\caption{Association Between Demographics and Eligibility Criteria and AHA Scores}
\label{table:demographics_on_score}
\end{table*}

\section{Alternative Specifications for RQ2}
\label{section:appendix_robustness_check}
In this section, we examine two alternative specification for RQ2 to illustrate the robustness of the results on the heterogeneous effects of AHA scores on service rate. 

The first alternative specification treats AHA Score as a continuous variable without segmenting assessments into different risk categories. In the second alternative specification, we create a binary variable \verb|high score| which equals to 1 if the AHA Score associated with the assessment is higher or equal to 9, and 0 otherwise. This allows us to more directly distinguish assessments with high risk scores from the rest. The results are shown in Table~\ref{table:robustness1} and Table~\ref{table:robustness2} respectively. 

Aligned with the result presented in the main text, we find the coefficients associated with the triple interaction terms of \verb|Black * Post * AHA Score| and \verb|Black * Post * High Score| to be not statistically significantly. This shows that the relationship between high AHA scores and service rates does not differ significantly by race in the post-intervention period, supporting the robustness of our findings. 

\begin{table*}[t]\centering
\def\sym#1{\ifmmode^{#1}\else\(^{#1}\)\fi}
\scalebox{0.85}
{
\begin{tabular}{l*{3}{c}}
\hline\hline
            &\multicolumn{1}{c}{(1)}&\multicolumn{1}{c}{(2)}&\multicolumn{1}{c}{(3)}\\
\hline
Black  &      0.2369         &      0.3413         &      0.3717         \\
            &    (0.2165)         &    (0.2628)         &    (0.2640)         \\
[1em]
Post        &     -1.6446\sym{***}&     -1.5083\sym{**} &     -1.6980\sym{**} \\
            &    (0.4610)         &    (0.5282)         &    (0.5580)         \\
[1em]
AHA Score       &      0.0824\sym{***}&      0.0191         &      0.0243         \\
            &    (0.0203)         &    (0.0250)         &    (0.0250)         \\
[1em]
Black * AHA Score &     -0.0714\sym{*}  &     -0.0776\sym{*}  &     -0.0830\sym{*}  \\
            &    (0.0306)         &    (0.0370)         &    (0.0372)         \\
[1em]
Post * AHA Score &      0.2246\sym{***}&      0.2324\sym{***}&      0.2244\sym{***}\\
            &    (0.0551)         &    (0.0595)         &    (0.0602)         \\
[1em]
Black * Post &     -0.7790         &     -1.1041         &     -1.2179         \\
            &    (0.6583)         &    (0.7398)         &    (0.7500)         \\
[1em]
Black * Post * AHA Score &      0.1256         &      0.1570         &      0.1717         \\
            &    (0.0787)         &    (0.0890)         &    (0.0901)         \\
\hline
N           &   6,143         &   6,143         &   6,143         \\
Pseudo. R\(^2\)      &      0.0282         &      0.2803         &      0.2830         \\
\hline\hline
\multicolumn{4}{l}{\footnotesize Standard errors in parentheses}\\
\multicolumn{4}{l}{\footnotesize \sym{*} \(p<0.05\), \sym{**} \(p<0.01\), \sym{***} \(p<0.001\)}\\
\end{tabular}
}
\caption{Service Rate by Race Risk Group, and AHA Implementation (Alternative Specification 1)}
\label{table:robustness1}
\end{table*}

\begin{table*}[t]\centering
\def\sym#1{\ifmmode^{#1}\else\(^{#1}\)\fi}
\scalebox{0.85}
{
\begin{tabular}{l*{3}{c}}
\hline\hline
            &\multicolumn{1}{c}{(1)}&\multicolumn{1}{c}{(2)}&\multicolumn{1}{c}{(3)}\\
\hline
Black  &     -0.1350         &     -0.1100         &     -0.1031         \\
            &    (0.0979)         &    (0.1164)         &    (0.1173)         \\
[1em]
Post        &     -0.0922         &      0.0033         &     -0.2449         \\
            &    (0.1424)         &    (0.2570)         &    (0.3025)         \\
[1em]
High Score &      0.3959\sym{***}&      0.0155         &      0.0344         \\
            &    (0.1161)         &    (0.1420)         &    (0.1424)         \\
[1em]
Black * High Score &     -0.3033         &     -0.2035         &     -0.2425         \\
            &    (0.1757)         &    (0.2123)         &    (0.2140)         \\
[1em]
Post * High Score  &      0.5890\sym{**} &      0.7660\sym{**} &      0.7443\sym{**} \\
            &    (0.2064)         &    (0.2437)         &    (0.2442)         \\
[1em]
Black * Post &     -0.0158         &     -0.0683         &     -0.0980         \\
            &    (0.2011)         &    (0.2376)         &    (0.2394)         \\
[1em]
Black * Post * High Score &      0.4139         &      0.3610         &      0.4240         \\
            &    (0.2932)         &    (0.3481)         &    (0.3501)         \\
\hline
N           &   6,143         &    6,143          &    6,143          \\
Pseudo. R\(^2\)       &      0.0210         &      0.2754         &      0.2782         \\
\hline\hline
\multicolumn{4}{l}{\footnotesize Standard errors in parentheses}\\
\multicolumn{4}{l}{\footnotesize \sym{*} \(p<0.05\), \sym{**} \(p<0.01\), \sym{***} \(p<0.001\)}\\
\end{tabular}
}
\caption{Service Rate by Race Risk Group, and AHA Implementation (Alternative Specification 2)}
\label{table:robustness2}
\end{table*}

\end{document}